\documentclass[10pt,conference]{IEEEtran}

\usepackage{amsmath}
\usepackage{balance}
\usepackage[noadjust]{cite}
\usepackage{mathptmx}
\usepackage{graphicx}
\usepackage{url}
\usepackage{array}
\usepackage{float}
\usepackage{booktabs} % For formal tables
\usepackage[utf8]{inputenc}
\usepackage{multirow}
\usepackage{colortbl}
\usepackage{tikz}
\usepackage{tabto}
\usepackage{enumitem}
\usepackage{booktabs}
\usepackage[normalem]{ulem}
\useunder{\uline}{\ul}{}

\usepackage{xargs}
\usepackage{xcolor}

\usepackage[hang]{footmisc}
\usepackage{tikz}

\usepackage{pifont}

\usepackage{hyperref}
\hypersetup{
	colorlinks=true,
	urlcolor=black,
	linkcolor=black,
	citecolor=blue,
	pdfstartview={FitH},
}

\begin{document}

\title{Assessing Simulation Knowledge and Proficiency Among Undergraduate Computing Students in Brazil: Insights and Results from a Survey Research}

\makeatletter
\newcommand{\linebreakand}{%
  \end{@IEEEauthorhalign}
  \hfill\mbox{}\par
  \mbox{}\hfill\begin{@IEEEauthorhalign}
}
\makeatother

\author{\IEEEauthorblockN{Fernando Brito Rodrigues, Valdemar Vicente Graciano Neto}
\IEEEauthorblockA{\textit{Federal University of Goiás (UFG)} \\
Goiânia, Brazil \\
\{valdemarneto\}@ufg.br, \\ \{fernandobrito\}@discente.ufg.br}
}

\frenchspacing
\maketitle

\begin{abstract} 
This paper reports results of an investigation about the level of knowledge among undergraduate computer science students in Brazil regarding the topic of simulation. Amid rapid technological evolution, simulation emerges as a crucial resource for training professionals capable of facing complex challenges. The research seeks to analyze the presence and effectiveness of simulation education, exploring students' perceptions, the tools used, the challenges faced, and the prospects for deeper study. This report highlights the importance of academic training in a dynamic technological environment, emphasizing the crucial role of simulation education in undergraduate computer science, while exploring the foundations of the methodologies and educational strategies associated with the topic. A survey research approach is adopted. 108 answers were received from 10 Brazilian states. 19 respondents from 15 different institutions said they had some contact with simulation during their studies. Results reveal that MATLAB/Simulink is the most popular formalism/tool used to teach simulation in Brazil.
\end{abstract}


\section{Introduction}
\label{sec:introduction}

In the dynamic and fast-paced context of contemporary technologies, academic training plays a crucial role in preparing professionals capable of facing the complex challenges of the real world. In the context of undergraduate computing courses in Brazil, the growing use of simulation tools has emerged as an essential component to enhance students' educational experience, providing a practical and innovative approach to learning.

The survey was motivated by the significant importance of simulation knowledge for undergraduate computing students in Brazil, which is a fundamental technique in several areas, providing the practical application of theories and concepts learned in the classroom. A growing concern is centered on the way simulation has been implemented in undergraduate computing courses in the country.

This monograph sought to explore the presence and effectiveness of simulation teaching in higher education institutions in Brazil, with a focus on undergraduate computing areas. The research aimed to provide a comprehensive understanding of the current state of simulation teaching, identifying both the opportunities and the challenges faced by students in this context.

The effectiveness of teaching through simulation is vital to the quality of training of future professionals in the area. The research sought to understand students' perceptions and level of knowledge regarding simulation, identifying gaps and challenges in its implementation. The objective is to outline guidelines that contribute to improving the integration of simulation in teaching, aligning academic training with the practical demands of the sector.

The research protocol was developed based on the guidelines proposed by \cite{kasunic:survey} Kasunic, \cite{Linaker:survey:guideline} Linaker and \cite{Molleri:2016:SGS:2961111.2962619} Molleri, by establishing a method structured into six stages: Introduction; Theoretical Foundation; Planning; Execution; Results and Conclusion.

Throughout this document, detailed analyses of students' perceptions and experiences regarding simulation teaching were presented, highlighting the most used tools, the challenges faced, and the prospects for further studies. The conclusions derived from these analyses will contribute to the development of recommendations aimed at improving the quality of simulation teaching in higher education institutions, thus consolidating the importance of this practice as a vital component in the training of computing professionals.

In addressing the challenges and opportunities in integrating simulation into teaching, the study reveals the heterogeneity in student responses, pointing to significant challenges such as the lack of updated content, financial inaccessibility of specific tools, and technical difficulties. These difficulties offer opportunities for improvement, highlighting the need for differentiated strategies to stimulate student engagement.

The research plan is detailed, outlining objectives, identification of the target audience, questionnaire design, definition of topics, evaluation scale, representative sampling, and informed consent. This plan aims to investigate the level of knowledge of undergraduate computing students in Brazil on the topic of simulation.

The questionnaire was presented, divided into sections such as Demographic Questions, Specific Questions, and Research Questions, each contributing to an in-depth understanding of the current state of simulation teaching. The questionnaire structure ranges from demographic questions to specific aspects of knowledge, practical application, tools used, learning sources, challenges faced and prospects for improving teaching.

By addressing specific issues such as basic simulation knowledge, tools and methods used, practical applications, perceived importance in training and learning sources, the questionnaire aims to capture a comprehensive view of students’ experience of simulation education.

In summary, this study not only provides an in-depth understanding of the current state of simulation education, but also offers valuable contributions to the development of more effective educational strategies. The conclusions and recommendations derived from the analysis of the data provide a guide for improving the quality of teaching, aligning it with contemporary demands and preparing students more effectively for the practical and professional challenges in their fields.

The remainder of the paper is organized as follows: Section 2 presents a brief background, Section 3 details the planning, conduction, and results of our study. Section 4 discusses the results. Finally, Section 5 concludes the paper.

\section{Background}
\label{sec:background}

Simulation emerges as a crucial pedagogical tool, allowing students a practical and dynamic approach to understanding complex concepts and real-world scenarios \cite{chwif2006modelagem}. In this context, the topic explores the fundamentals of simulation and simulation teaching.
\\

Simulation is a controlled technique that replicates essential aspects of the real world in simulated environments. It involves the creation of representative models, which can be physical, mathematical, statistical or computational, depending on the phenomenon to be simulated. This approach offers the ability to control variables and experimental conditions \cite{reichard1992computer}, and specifically in computing courses, simulation is considered one of the most widely used research methods to conduct exploratory empirical investigations \cite{Molleri:2016:SGS:2961111.2962619}. Amid rapid technological evolution, it emerges as a crucial resource for training professionals capable of facing complex challenges. \\

To illustrate the simulation, we can demonstrate the use of one of the most widely used simulation tools in current research, which is the Matlab tool. Matlab is widely recognized for its ability to simulate complex systems accurately and efficiently. By using Matlab to simulate the operation of a wind turbine, we can model all of its physical components and analyze its performance under different conditions.

Simulating a wind turbine in Matlab allows us to explore a wide range of variables, from the different scenarios and environmental conditions that the turbine may face to the specific characteristics of its physical components, such as the blades, the generator and the control system. In addition, it is possible to simulate variations in voltage, electric current and other electrical characteristics of the turbine, enabling a detailed analysis of its behavior in different situations, as shown in the image below. \cite{souza2022simulador}.

Modeling allows researchers to isolate specific factors and observe their impact on the system as a whole, and in environments where direct experimentation is dangerous, impractical or too expensive, simulation offers a safe and affordable alternative. For example, in aviation, flight simulators allow pilots to be trained in emergency situations without real risks, and this is a case where "not simulating" can cause considerable costs, such as wrong assumptions or expectations or even risk to life \cite{DBLP:conf/ispw/Birkholzer12}. In addition, simulation generates data that can be analyzed to better understand the behavior of the system under study, leading to process improvements, resource optimization or more informed decision-making.

To complement the theoretical foundation, simulation education refers to the use of simulation as an educational tool. This approach provides students with the opportunity to apply theories and concepts learned in the classroom in practical and realistic contexts \cite {bradley2014review}. Some characteristics of simulation education include:
\noindent \textbf{Hands-on Experience:} Simulation provides a hands-on experience that goes beyond theory, allowing students to develop practical skills and acquire contextualized knowledge \cite{weis1998computer};
\noindent \textbf{Decision Making:} Simulated environments often present challenges that require quick and effective decision-making \cite {garrett2001value}. This helps develop critical thinking and problem-solving skills; \noindent \textbf{Immediate Feedback:} Simulation allows for immediate feedback on actions and decisions \cite{tena2017training}, enabling error correction and continuous improvement of performance;
\noindent \textbf{Controlled Environments:} Educators can create controlled environments to simulate specific situations, ensuring that students face challenges relevant to the field of study;
\noindent \textbf{Interdisciplinary Application:} Simulation is an approach that can be applied in various disciplines, from healthcare (medical simulation) to engineering, management and military training \cite{harder2010use}.
\section{Survey Research}
\label{sec:slr}

Survey research is a technique used to acquire knowledge by listening to students' voices and observing their responses, attitudes, and behaviors in order to gain a broader understanding of a given topic of interest. In this context, the objective of the study is to investigate the level of knowledge of undergraduate computer science students in Brazil on the topic of simulation. The purpose of the research is to understand how students perceive the use of simulation in practice. A survey protocol and corresponding questionnaire were developed to collect perceptions of computer science students on the use of simulation through an online questionnaire.

To identify the target audience, the survey was composed of undergraduate computer science students from higher education institutions throughout Brazil. The questionnaire consisted of questions that addressed different aspects of simulation, such as basic concepts, simulation methods, applications, and practical knowledge, as well as how simulation is taught (or not) at the respondents’ institutions, and the questions were designed to cover both theoretical and practical aspects.

Based on the literature review, the main topics covered in the questionnaire were defined. This included basic concepts of simulation, types of simulation, applications in different areas, simulation tools, and practical simulation cases. Each topic was divided into specific questions to assess students' knowledge regarding these topics.

A rating scale for the closed questions in the questionnaire was defined, allowing students to indicate their level of knowledge in each topic. This allowed quantifying the degree of students' knowledge in different areas of simulation.

A representative sample of undergraduate students from different higher education institutions in Brazil was carried out. For this purpose, several universities and undergraduate courses will be contacted to ensure the geographical diversity of participants. Before starting the survey, students will be provided with detailed information about the objectives of the survey, the procedures involved, and data privacy. Students will be asked to provide their informed consent before participating in the survey.

\subsection{Form}
 The structure of the questionnaire was designed as illustrated in Table 2.1. We developed (i) demographic questions (DQ), that is, characterization questions to obtain the profile of the participants, (ii) specific questions (QE) to ask the participants specific information about the topic, and we linked the research questions (QP), whose purpose was to collect the perceptions of the participants after solving the questions about the proposed topic. \\

Groups of closed questions (QF) and open questions (QA) were defined. Closed questions present predefined options from the respondents, while open questions allow you to provide detailed answers.

% \usepackage{multirow}
\begin{table}[H]
\caption{Questionnaire Layout}
\label{cap:descr}
\centering
\begin{tabular}{|l|l|}
\hline
\textbf{Demographic Questions} & Characterization Questions \\ \hline
\multirow{2}{*}{\textbf{Specific Questions}} & Closed-Ended Questions \\ \cline{2-2}
& Open-Ended Questions \\ \hline
\textbf{Research Questions} & Subquestions \\ \hline
\end{tabular}
\end{table}

\noindent \textbf {Research Questions:}
\label{subsec:research-planning}

To answer the demographic questions, research, students' perceptions, experiences and interactions they have had about teaching simulation in Brazil were collected, as well as what they perceive as positive and negative on the topic, through specific questions. To achieve the established research objectives, therefore, the following research questions were derived (Table 2.2), and their relationship with the specific questions (Table 2.3):

% \usepackage{booktabs}
% \usepackage{graphicx}
\begin{table}[H]
\centering
\caption{Research Questions}
\label{cap:descr}
%\resizebox{\textwidth}{!}{%
\begin{tabular}{@{}p{2cm}p{6cm}@{}}
\toprule
\textbf{ID} & \textbf{Research Questions (RQ) and Research Subquestions (SubRQ)} \\ \midrule
RQ1 & How is simulation teaching carried out in Brazil? \\ \midrule
SubRQ2 & In which institutions? \\ \midrule
SubQP3 & In which courses, level or context (undergraduate, postgraduate, extension, research project) and at what point (in the course) do they learn simulation? \\ \midrule
SubQP4 & What tools/notations are taught in the course in which the respondent is enrolled? \\\hline
\end{tabular}
\end{table}

\subsection{Conduction}

Before the questionnaire was applied to the target audience, a pilot study was conducted with a small sample of students to identify potential problems or adjustments needed in the questionnaire. This small sample made it possible to change the form before the questionnaire was sent to the other students, and contributed to the quality of the questions.

The questionnaire was sent to undergraduate students from several higher education institutions in Brazil via email and Google forms, and access to the questionnaire was protected by a password or exclusive link to ensure the confidentiality of the responses.

External monitoring of the students was carried out to encourage participation and to send reminders to those who did not respond to the questionnaire. This helped to increase the response rate and obtain a more representative sample. Internal monitoring was also carried out, where the deadlines for the survey were maintained in accordance with the stages and reminders for completing the activities.

After data collection, a quantitative analysis of the results of the closed-ended questions was performed using statistical techniques. In addition, a qualitative analysis of the open-ended responses was performed, identifying emerging themes and patterns related to the students' knowledge about simulation.

The research results were interpreted and discussed in light of the relevant literature on simulation and higher education. Comparisons were made with similar studies in other countries to contextualize the results.

\subsection{Results}

This report analyzed the perceptions of undergraduate students in higher education in computing in Brazil about simulation education in the country, based on responses collected through a survey. Divided into sections, the report addressed specific issues, from identifying institutions that offer simulation education to assessing the quality of education and the importance perceived by students. In addition, the report highlighted the simulation tools and methods used, the challenges faced by students and their suggestions for improvements. The analysis also addressed basic simulation knowledge, participation in academic and extracurricular projects, and students' interest in furthering their studies after graduation.

By providing a detailed view of students' experiences and perceptions, this report aimed to provide insights to improve simulation education in the Brazilian context. Conclusions and recommendations were presented based on the research questions, aiming to contribute to the development of more effective educational strategies, addressing specific challenges and promoting a more enriching learning environment aligned with current demands.

In total, 108 responses were obtained. However, 76 participants indicated 'No' to the 'Elimination Question', excluding them from the remaining questions due to lack of knowledge or involvement with the simulation topic. Subsequent analyses are based on the responses of the remaining 32 students who had contact with simulation, 4 women and 28 men. The research covers 10 different states, namely: Minas Gerais, Pará, Mato Grosso, Maranhão, Ceará, Goiás, São Paulo, Rio de Janeiro, Rio Grande do Sul and Paraná, contributing significantly to a comprehensive and representative overview of the sample collected through the questionnaire.

\noindent \textbf {QP1/SubQP2 – In which institutions do we have simulation teaching in Brazil?}

We obtained 32 responses about the institutions in which the public had some contact with the study of simulations, and only 19 of these responses confirmed that they had some partial study on the subject. The following is an analysis of the data, along with an image representing the states in which simulation teaching takes place, as indicated in the responses.

\begin{figure}[H]
\centering
\includegraphics[width=0.5\linewidth]{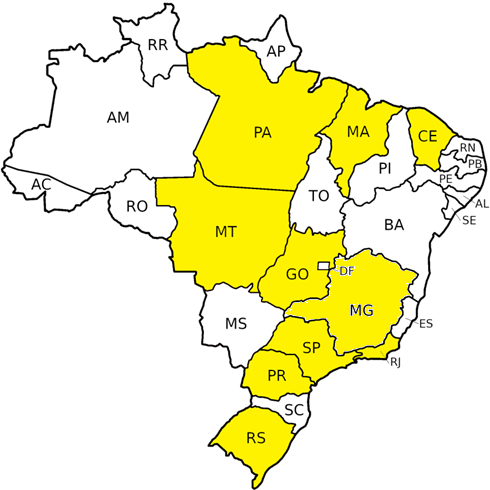}
\caption{Brazil: States in which there was contact with simulation teaching.}
\end{figure}

Three students from the Federal University of Maranhão (UFMA) were introduced to simulation through the "Control Engineering and Robotics" course. At the Federal University of Pará (UFPA), three students explored the topic during the "Discrete Simulation" course. In addition, at the Federal University of Goiás (UFG), two students participated in a summer school dedicated to simulation. At the Technological Institute of Aeronautics (ITA), three students conducted a partial simulation study during the "Control Systems" course. Three other students from the State University of Western Paraná (Unioeste) acquired knowledge about simulation through the Artificial Intelligence course.

At both the Federal University of Rio de Janeiro (UFRJ) and the Federal University of Ceará, one student from each institution became familiar with simulation in the ``Performance Analysis'' course. In addition, at the University of São Paulo (USP), a student explored simulation in the "Numerical Calculus" courses and during his scientific initiation. In turn, at the Federal Institute of Education, Science and Technology of Rio Grande do Sul (IFRS), a student confirmed having had contact with simulation. Finally, at the University of Cuiabá (UNIC), a student used simulation software in his master's degree course. These diverse experiences highlight the relevance of simulation in different academic contexts. \\\\
\noindent \textbf{QP1/SubQP3 - In which courses, level or context (undergraduate, postgraduate, extension, research project) and at what point (in the course) do they learn simulation?} \\

According to the students' responses, learning in simulation is related to subjects offered in the undergraduate courses, but there are some cases in which the study of simulation also happened through other options. The data analysis is as follows:

\textbf{Number of students per course:}
\begin{itemize}
\item {Computer Science} - 15 Students;
\item {Computer Engineering} - 9 Students;
\item {Information Systems} - 5 Students;
\item {Systems Analysis and Development} - 2 Students; \item {Software Analysis and Development} - 1 Student;
\end{itemize}

\textbf{Number of students per university:}

\begin{itemize}
\item {Federal University of Goiás} - 4 Students;
\item {Federal University of Maranhão} - 4 Students;
\item {Western Paraná State University} - 4 Students;
\item {Technological Institute of Aeronautics} - 3 Students;
\item {Federal University of Pará} - 3 Students;
\item {University of São Paulo} - 2 Students;
\item {Federal University of Juiz de Fora} - 2 Students;
\item {Federal University of Ceará} - 2 Students;
\item {SENAI College of Technology} - 2 Students;
\item {Federal University of Rio de Janeiro} - 1 Student;
\item {Federal University of ABC} - 1 Student;
\item {Federal Institute of Rio Grande do Sul} - 1 Student; \item {State University of Campinas} - 1 Student;
\item {University of Cuiabá} - 1 Student;
\item {UEG De Santa Helena} - 1 Student.
\end{itemize}

The study revealed that students from several Brazilian universities have been actively involved in learning and applying simulation in different academic contexts. One example is a student from UFG, who acquired knowledge about simulation through participation in a summer school and Hackathon. In this event, sensors were used to monitor crops on a farm, providing a practical application of the concepts learned. At UFJF, two students explored simulation as part of their scientific initiation activities, expanding their understanding of this technique. At the Federal University of ABC, a master's student conducted a systematic mapping addressing the simulation of emergent behaviors in Systems-of-Systems (SoS). Meanwhile, a student from Unicamp dedicated himself to a scientific initiation focused on fluid simulation with a focus on computer graphics.

At UNIC, a student was involved in the assembly of a meteorological station through a postgraduate program, applying simulation to collect and analyze meteorological data. At USP, students participated in various activities, such as joining a robotics extension group, where one of them had the opportunity to explore simulation in a practical context. In addition, a student from USP dedicated himself to scientific initiation, delving deeper into the simulation of systems software architectures. Finally, at UEG in Santa Helena, a student had contact with simulation through a software development project, intended for use by a real company. These experiences highlight the diversity of approaches and applications of simulation in higher education institutions, providing students with a rich and comprehensive education in the field of virtual modeling. \\\\
\noindent \textbf{QP1/SubQP4 - What tools/notations are taught in the course in which the respondent is enrolled?} \\

The data analysis and tables below show which tools are taught in the course in which the respondent is enrolled, according to the students' responses:

\begin{figure}
[H]
\centering
\includegraphics[width=1\linewidth]{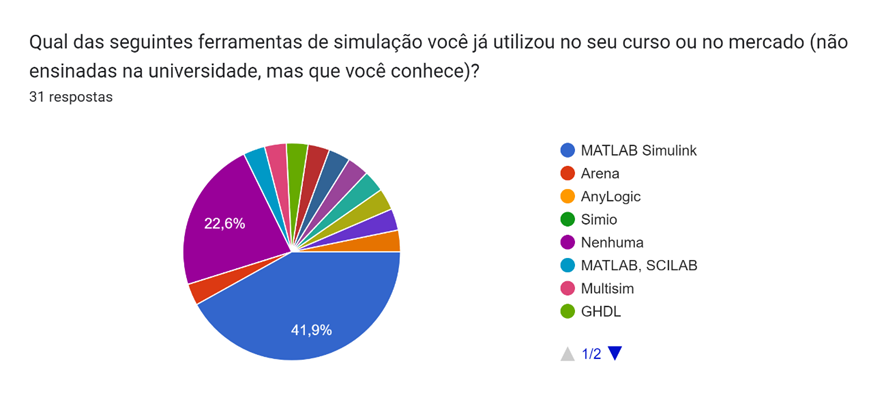}

\centering
\includegraphics[width=1\linewidth]{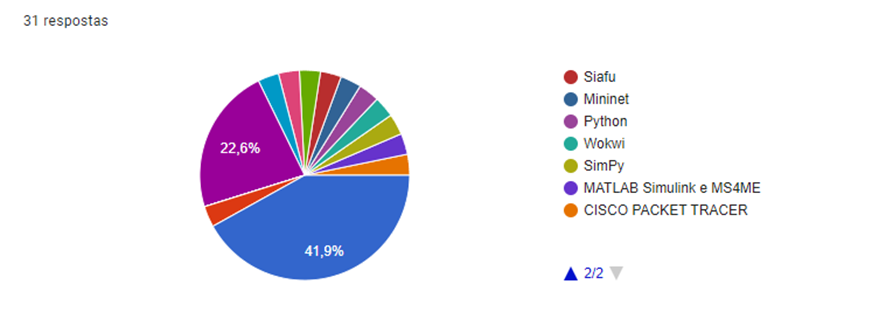}
\caption{Graph of tools used for simulation}

\end{figure}
The study revealed that the most widely used tool by participants is MATLAB Simulink, with a significant 41.9\% indicating its use. This predominance highlights the relevance of directing attention to this specific tool in the context of teaching simulation. However, most students did not mention using specific tools such as Arena, AnyLogic or Simio, suggesting a possible gap in exposure to these platforms. A significant number of respondents, corresponding to 22.6\%, indicated that they had not used any simulation tools outside of the academic environment. In addition to MATLAB and Simulink, other tools were mentioned, such as MATLAB and SCILAB, Multisim, GHDL, Siafu, Mininet, Python, Wokwi, SimPy and CISCO PACKET TRACER, but were reported by a smaller number of participants. This diversity in tool use highlights the variety of approaches and technologies available in the simulation field, highlighting areas that may require greater emphasis in teaching and academic practice. 

\begin{figure}
[H]
\centering
\includegraphics[width=1\linewidth]{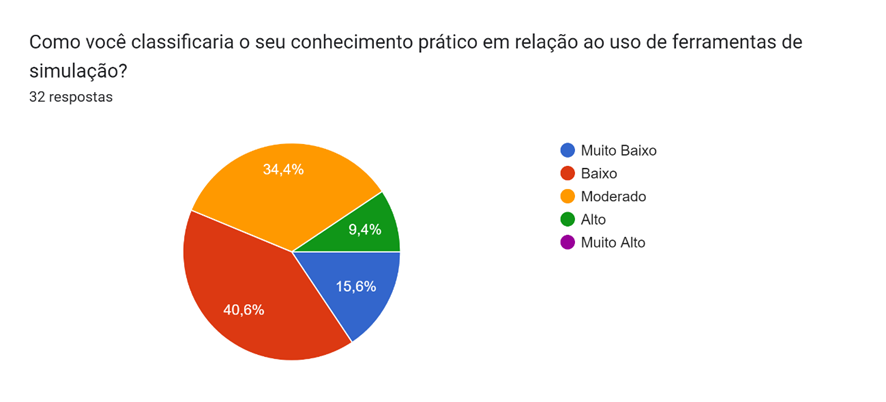}
\caption{Graph showing the level of knowledge of simulation tools}

\end{figure}

The assessment of practical knowledge among the participants presented a variety of classifications, with the majority indicating levels considered as Low (40.6\%) or Moderate (34.4\%). Notably, a small contingent of respondents revealed having a knowledge categorized as Very Low (15.6\%), while a smaller portion indicated having an understanding considered High (9.4\%). No participant classified their knowledge as Very High. This diversity in perceptions reflects the heterogeneity in the level of familiarity and practical experience of the participants in the field of simulation.
\\\\
\noindent \textbf{QP1/SubQP5 - What difficulties did you encounter and how do you envision improving simulation teaching?} \\

The students who participated in the survey highlighted several difficulties in studying simulation. Many pointed out the lack of updated content as a challenge, highlighting the obsolescence of available materials and the incompatibility of old tools. The financial inaccessibility of paid tools was also a common concern, limiting access for students. In addition, some participants faced technical challenges when dealing with simulation software, highlighting the complexity and difficulty of interaction.

The scarcity of educational materials and tutorials was mentioned as a barrier to effective learning, as was the lack of time, especially during periods of tight deadlines in the master's degree. The use of outdated hardware was also pointed out as a difficulty, affecting the efficiency in simulations. The lack of content directed to different areas of activity in simulations was identified as a gap in teaching.

Additionally, some students mentioned specific challenges, such as the complexity of probability distributions in queueing theory simulation. Difficulty in controlling the environment and interacting with the software, along with issues with simulator documentation, were cited as additional obstacles. The scarcity and difficulty in understanding available material, especially when paid or of questionable quality, were also points of concern.

These diverse experiences reflect the complexity of simulation study and highlight specific areas that could be targeted for improvement in teaching, support, and accessibility to educational resources.

\begin{figure}
[H]
\centering
\includegraphics[width=1\linewidth]{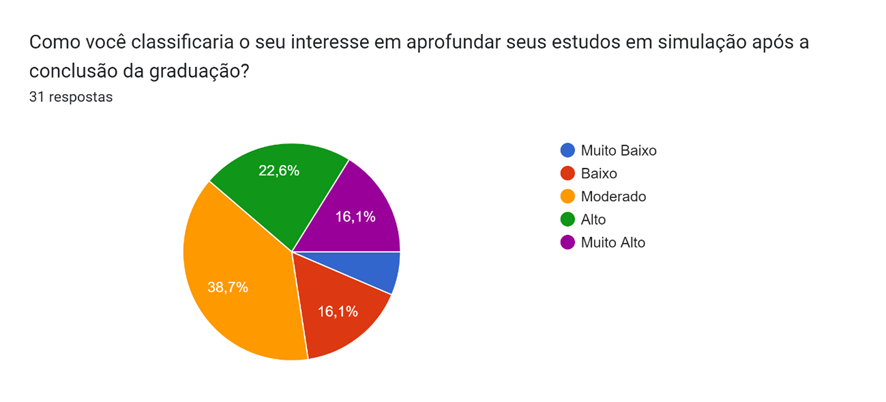}
\caption{Simulation Studies Interest Graph}

\end{figure}

Based on responses from 31 participants, the survey revealed significant diversity in students’ levels of interest in furthering their studies in simulation after graduation. The majority of participants, corresponding to 38.7\% of the total (12 students), expressed an interest classified as moderate. Notably, a significant percentage of 22.6\% (7 students) demonstrated high interest, while 16.1\% (5 students) indicated very high interest. This varied distribution highlights the breadth of students’ perspectives and dispositions toward pursuing simulation studies further after completing their undergraduate studies. Furthermore, the identification of these different levels of interest suggests opportunities and challenges in teaching simulation. The presence of a significant group with high or very high interest suggests the opportunity to develop specific programs, such as postgraduate courses, workshops, or research projects, to meet the academic aspirations of these students. On the other hand, the moderate or low interest of a significant portion of the students highlights the importance of stimulation strategies, such as hands-on activities, relevant case studies, and events that can spark greater engagement and passion for simulation. Regarding the classification of practical knowledge, the data reveal a variety of levels among the participants. The majority classified their knowledge as Low (40.6\%) or Moderate (34.4\%). A small number indicated that they had Very Low knowledge (15.6\%), and 9.4\% rated their knowledge as High. Surprisingly, no participants rated their knowledge as Very High. This heterogeneity highlights the need for flexible and personalized educational approaches to meet students' different demands and levels of familiarity with simulation practice.
\\\\
\\\\
\noindent \textbf{QP1/SubQP 6 - How do students judge the quality of simulation teaching?} \\

Students evaluated the incidence of simulation teaching in their institutions, revealing a diversity of perceptions:

\begin{figure}
[H]
\centering
\includegraphics[width=1\linewidth]{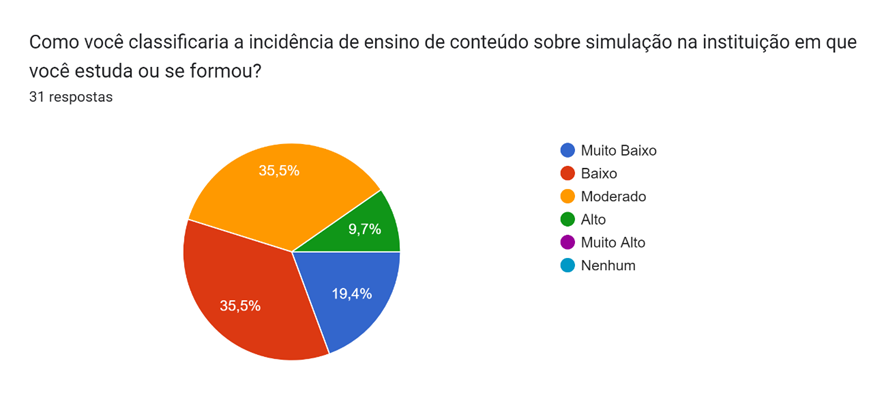}
\caption{Graph showing the incidence of simulation teaching in institutions}

\end{figure}

Of the 31 participants, the majority indicated a low (35.5\%) or very low (19.4\%) incidence, suggesting a possible gap in the coverage of simulation content in the curricula, with 11 students reporting low incidence and about 6 students mentioning a very low incidence.

A significant portion, equivalent to 35.5\% (11 students), classified the incidence as moderate, indicating an intermediate level of exposure to simulation teaching. On the other hand, a minority, representing 9.7\% of students (approximately 3 students), perceived a high incidence of simulation education at their institutions.

Notably, 19.4\% (6 students) reported no incidence of simulation education at their institutions, highlighting a significant absence of this component in educational curricula. These varied perceptions highlight the need for further analysis on the integration of simulation education in educational institutions.

These results indicate challenges in effectively integrating simulation into education, highlighting the need for improvements and expansion of simulation education in institutions. The diversity in responses suggests variation in educational approaches, highlighting the importance of more consistent integration into curricula. The absence of teaching for some students suggests the need for institutional assessment and strategies to include this valuable tool in the educational environment.

\subsubsection*{1.15/QA1 - In your opinion, how important is simulation as a learning tool in your undergraduate field?}

We obtained 29 responses, where students highlighted the significant importance of simulation in their undergraduate fields. For many, simulation is a fundamental tool for teaching real-world processes, offering a valuable alternative when direct experience is challenging. In specific areas, such as hardware, automation and industrial automation, where direct access to tools is limited, simulation was considered crucial.

In addition, simulation was recognized as an essential preparation for the job market, allowing practice in a controlled environment. The reduction of costs and losses, the contribution to the country's development and the ability to visualize projects in early stages were highlighted as benefits.

Simulation was also perceived as a valuable tool for deepening the understanding of theoretical concepts, providing their application in real-world scenarios. In areas such as robotics, artificial intelligence and software engineering, simulation played a crucial role, contributing to a better understanding of complex systems.

In summary, students see simulation as an essential tool for enhancing learning, offering practical and applied benefits across a variety of disciplines. These perceptions highlight the importance of effectively integrating simulation into teaching, aiming to provide an educational experience that is more aligned with market demands and the practical needs of students.

\section{Threats to Validity and Limitations of the Research}
\label{sec:discussion}

By exploring the results, challenges and critical points that may influence the reliability and generalizability of the conclusions were identified, aiming to provide a critical reflection on the research conducted, highlighting areas for improvement and important considerations for future research.

The study examined the perceptions of undergraduate computer science students about simulation teaching in Brazil, aiming to provide \textit{insights} to improve educational strategies. However, it is crucial to consider the potential threats to validity and limitations that may impact the interpretation of the results.

Data collection was based on questionnaire responses, which are subject to individual interpretation and possible response bias. In addition, the exclusion of participants who answered negatively to the "Elimination Question" may have influenced the representativeness of the sample.

The generalizability of the results to the entire country may be affected by the concentration of responses in certain regions and institutions. Geographic diversity may not have been fully addressed, compromising external validity.

The definition of research questions and the choice of variables may have limited the scope of the study, influencing construct validity. Differences in the interpretation of key terms by participants may have impacted the consistency of the data.

The sample size, especially after excluding participants, may have impacted the statistical validity of the analyses. Significant results may be influenced by the heterogeneity of the sample.

In addition, the questionnaire used has limitations that should be considered. The approach adopted may have restricted the depth of responses, failing to capture important nuances in the students' experiences. The nature of the closed questions may not have provided a complete understanding of the complexity of the topic in question.

Another limitation identified refers to the concentration of responses in certain states, which may not adequately represent the geographic diversity of Brazil. The inclusion of more regions could enrich the analysis and provide a more comprehensive view of the educational landscape in relation to teaching simulation in undergraduate computing courses.

These limitations highlight the importance of interpreting the results with caution and recognizing the restrictions inherent in the methodology adopted. Future research could explore more comprehensive approaches, considering the inclusion of participants from different regions and using methods that allow for a deeper understanding of students’ experiences in simulation education.
\section{Final Remarks and Future Work} 
\label{sec:FR}
This study provided a comprehensive analysis of simulation teaching in undergraduate computer science courses in Brazil, highlighting the importance of this topic in students’ academic training. Throughout the research, several aspects were addressed, from the methodology used to the results obtained, with the aim of offering insights to improve teaching and understand students’ perceptions.

The research highlighted the growing importance of simulation teaching in students’ training, recognizing it as a fundamental tool for teaching real-world market processes, especially when direct experience is challenging. The diversity of tools used by students, with emphasis on MATLAB Simulink, highlights the variety of approaches and technologies available in this field, although the lack of exposure to some tools suggests possible gaps in the integration of these platforms into educational curricula.

The challenges identified, such as lack of updated content, financial inaccessibility of paid tools, and technical difficulties, highlight the complexity of simulation studies. The scarcity of educational materials and tutorials is highlighted as a significant barrier to effective learning. Difficulties and challenges are encountered in the teaching of information systems in Brazil, including from the perspective of teachers \cite{Neves2023}.

The diversity in the levels of interest of students in deepening their studies in simulation after graduation highlights opportunities and challenges in teaching this practice. The assessment of practical knowledge reveals heterogeneity among participants, pointing to the need for flexible educational approaches.

The analysis of the incidence of simulation teaching in institutions revealed a variety of perceptions, with some students indicating low or no incidence. This diversity highlights the importance of a more in-depth analysis on the integration of simulation teaching in the curricula of higher education institutions.

The recognition of simulation as an essential tool to improve learning, offering practical and applied benefits in several disciplines, stood out in the conclusions. Recommendations for improving simulation teaching include updating content, expanding the variety of tools taught, and developing strategies to overcome identified challenges.

The study contributed to the development of more effective educational strategies, addressing specific challenges and promoting a more enriching learning environment aligned with contemporary demands. Despite the limitations identified, the work provided valuable information about the presence of simulation in educational institutions, the tools used, the difficulties faced by students, and the interest in further studies in this area after graduation. These conclusions can serve as a basis for improvements in teaching and future research directions.
\section{Acknowledgements}
\label{sec:ack}

XXXX

\bibliographystyle{IEEEtran}
\typeout{}
\bibliography{bib}

\begin{thebibliography}{10}
\providecommand{\url}[1]{#1}
\csname url@samestyle\endcsname
\providecommand{\newblock}{\relax}
\providecommand{\bibinfo}[2]{#2}
\providecommand{\BIBentrySTDinterwordspacing}{\spaceskip=0pt\relax}
\providecommand{\BIBentryALTinterwordstretchfactor}{4}
\providecommand{\BIBentryALTinterwordspacing}{\spaceskip=\fontdimen2\font plus
\BIBentryALTinterwordstretchfactor\fontdimen3\font minus \fontdimen4\font\relax}
\providecommand{\BIBforeignlanguage}[2]{{%
\expandafter\ifx\csname l@#1\endcsname\relax
\typeout{** WARNING: IEEEtran.bst: No hyphenation pattern has been}%
\typeout{** loaded for the language `#1'. Using the pattern for}%
\typeout{** the default language instead.}%
\else
\language=\csname l@#1\endcsname
\fi
#2}}
\providecommand{\BIBdecl}{\relax}
\BIBdecl

\bibitem{kasunic:survey}
M.~Kasunic, ``Designing an effective survey,'' Carnegie Mellon Software Engineering Institute, Pittsburg, USA, Tech. Rep. CMU/SEI-2005-HB-004, 01 2005.

\bibitem{Linaker:survey:guideline}
J.~Linåker, S.~Sulaman, M.~Host, and R.~de~Mello, ``Guidelines for conducting surveys in software engineering,'' Lund University, Sweden, Tech. Rep., May 2015.

\bibitem{Molleri:2016:SGS:2961111.2962619}
J.~S. Moll{\'e}ri, K.~Petersen, and E.~Mendes, ``Survey guidelines in software engineering: An annotated review,'' in \emph{10th ESEM}, Ciudad Real, Spain, 2016.

\bibitem{chwif2006modelagem}
L.~Chwif and A.~C. Medina, \emph{Modelagem e simula{\c{c}}{\~a}o de eventos discretos}.\hskip 1em plus 0.5em minus 0.4em\relax Afonso C. Medina, 2006.

\bibitem{reichard1992computer}
D.~Reichard, H.~Zhu, R.~Fox, and R.~Brazee, ``Computer simulation of variables that influence spray drift,'' \emph{Transactions of the ASAE}, vol.~35, no.~5, pp. 1401--1407, 1992.

\bibitem{souza2022simulador}
P.~H. B.~d. Souza \emph{et~al.}, ``Simulador did{\'a}tico para estudo de controle de pot{\^e}ncia de turbinas e{\'o}licas,'' 2022.

\bibitem{DBLP:conf/ispw/Birkholzer12}
\BIBentryALTinterwordspacing
T.~Birkh{\"{o}}lzer, ``Software process simulation is simulation too - what can be learned from other domains of simulation?'' in \emph{2012 International Conference on Software and System Process, {ICSSP} 2012, Zurich, Switzerland, June 2-3, 2012}, D.~R. Jeffery, D.~Raffo, O.~Armbrust, and L.~Huang, Eds.\hskip 1em plus 0.5em minus 0.4em\relax {IEEE} Computer Society, 2012, pp. 223--225. [Online]. Available: \url{https://doi.org/10.1109/ICSSP.2012.6225972}
\BIBentrySTDinterwordspacing

\bibitem{bradley2014review}
E.~G. Bradley and B.~Kendall, ``A review of computer simulations in teacher education,'' \emph{Journal of Educational Technology Systems}, vol.~43, no.~1, pp. 3--12, 2014.

\bibitem{weis1998computer}
P.~A. Weis and J.~Guyton-Simmons, ``A computer simulation for teaching critical thinking skills,'' \emph{Nurse Educator}, vol.~23, no.~2, pp. 30--33, 1998.

\bibitem{garrett2001value}
B.~M. Garrett and D.~Callear, ``The value of intelligent multimedia simulation for teaching clinical decision-making skills,'' \emph{Nurse Education Today}, vol.~21, no.~5, pp. 382--390, 2001.

\bibitem{tena2017training}
F.~Tena-Chollet, J.~Tixier, A.~Dandrieux, and P.~Slangen, ``Training decision-makers: Existing strategies for natural and technological crisis management and specifications of an improved simulation-based tool,'' \emph{Safety science}, vol.~97, pp. 144--153, 2017.

\bibitem{harder2010use}
B.~N. Harder, ``Use of simulation in teaching and learning in health sciences: A systematic review,'' \emph{Journal of Nursing Education}, vol.~49, no.~1, pp. 23--28, 2010.

\bibitem{Neves2023}
\BIBentryALTinterwordspacing
V.~de~Oliveira~Neves, S.~M. Melo, D.~Viana, R.~P. dos Santos, and V.~V.~G. Neto, ``Challenges on the brazilian information systems education: The professors' perspective,'' \emph{{IEEE} Trans. Educ.}, vol.~66, no.~6, pp. 531--542, 2023. [Online]. Available: \url{https://doi.org/10.1109/TE.2023.3259335}
\BIBentrySTDinterwordspacing

\end{thebibliography}

\end{document}